\documentclass[runningheads]{llncs}
\usepackage{appendix}
\usepackage[table]{xcolor}
\usepackage{graphicx}
\usepackage{subcaption} 
\usepackage{amsmath} 
\usepackage{comment}
\usepackage{multirow}
\usepackage[hidelinks]{hyperref}
\usepackage{url}

\usepackage{array, caption, tabularx,  ragged2e,  booktabs}
\usepackage{todonotes}
\usepackage{color,soul}
\begin{document}
%
\title{Exploring the Nexus Between Retrievability and \\ Query
  Generation Strategies\protect\footnote{Accepted at ECIR 2024.}}
\titlerunning{Nexus Between Retrievability and Query Generation Strategies}
%
%
\author{Aman Sinha$^\dagger$ 
\and
Priyanshu Raj Mall$^\dagger$ 
\and
Dwaipayan Roy$^*$ 
}


%



\authorrunning{A. Sinha et al.}
%
\institute{Indian Institute of Science Education and Research Kolkata, India\\
\email{\{as18ms065, prm18ms118, dwaipayan.roy\}@iiserkol.ac.in}\\
$^\dagger$These authors contributed equally to this work.\\
$^*$Corresponding author
}
\maketitle              
\begin{abstract}

Quantifying bias in retrieval functions through document retrievability scores is vital for assessing recall-oriented retrieval systems. However, many studies investigating retrieval model bias lack validation of their query generation methods as accurate representations of retrievability for real users and their queries. This limitation results from the absence of established criteria for query generation in retrievability assessments. Typically, researchers resort to using frequent collocations from document corpora when no query log is available. 
In this study, we address the issue of reproducibility and seek to validate query generation methods by comparing retrievability scores generated from artificially generated queries to those derived from query logs. Our findings demonstrate a minimal or negligible correlation between retrievability scores from artificial queries and those from query logs. This suggests that artificially generated queries may not accurately reflect retrievability scores as derived from query logs. We further explore alternative query generation techniques, uncovering a variation that exhibits the highest correlation. This alternative approach holds promise for improving reproducibility when query logs are unavailable.

\keywords{Reproducibility  \and Retrievability \and Empirical Study.}
\end{abstract}

\section{Introduction}\label{sec:intro}

In the field of Information Retrieval (IR), the study of biases and fairness of retrieval models together with their effectiveness has been an active area of research~\cite{ludovico2023bias,ekstrand2022foundations}.
Among the various metrics and methodologies developed,
\emph{retrievability} has emerged as a significant way of measuring 
%
how accessible a document is in a collection, given a retrieval system and a set of queries~\cite{azzopardi2008retrievability}.
It reflects the degree to which the system favors certain documents while retrieval over others, regardless of their relevance. Retrievability can be used to evaluate the fairness, diversity, and coverage of retrieval systems, as well as to identify and analyze the factors that affect the accessibility of documents~\cite{azzopardi2008accessibility}.
However, calculating the retrievability of a collection of documents is a complex task comprising several interconnected steps. 
Initially, a set of queries is selected based on which the retrievals are performed.
Researchers have used query logs as well as artificially generated queries in this step.
Subsequently, a retrieval model is configured with optimal parameters and the actual retrieval process follows, producing ranked document lists. 
Retrievability scores are then computed based on the rank of the documents in the retrieved lists.
Finally, the computed scores are used to estimate ``retrievability bias'' by metrics like the Gini coefficient\cite{gini1936measure}, Atkinson index~\cite{atkinson}, Theli index~\cite{theli} or Palma index~\cite{palma} borrowed from economics, to quantify bias and inequality (mainly in wealth) within a population.

Various query generation techniques employed at the initial stage can yield distinct query sets, potentially resulting in diverse retrievability scores for the same collection and retrieval system. Previous research has indicated that retrievability scores calculated using simulated queries differ from those obtained using actual queries extracted from query logs~\cite{traub2016querylog}. However, a comprehensive and systematic comparison of retrievability scores computed using various query simulation methods has yet to be undertaken.
The discrepancy, if any, poses a challenge to the reproducibility of retrievability experiments as different researchers may employ different query generation methods, leading to diverse conclusions on the same collection. To investigate the potential impact on result reproducibility and experimental validation, this paper conducts a comparative analysis of computed retrievability scores using the most commonly adopted query generation techniques in the research community.
The main contributions of this paper include:
\begin{itemize}
\item Providing an overview of query generation techniques commonly used in retrievability experiments documented in the literature.
\item Conducting a comprehensive empirical assessment of the reproducibility of retrievability experiments employing different query generation techniques and test collections.
\end{itemize}
Additionally, we address the following research questions in this paper:
\begin{itemize}
\setlength{\leftmargin}{2cm} 
    \item[\textbf{RQ1 - }] While keeping the other parameters constant, does varying the query simulation techniques lead to differences in retrievability scores, thus impacting the reproducibility of results?
    \item[\textbf{RQ2 - }] Do these differences in retrievability scores exhibit consistency across various collections, thereby influencing the reproducibility of results?
\end{itemize}

The rest of the paper is organized as follows.
In the next section, we provide essential background information and review previous research related to retrievability, with a particular emphasis on the prevalent query simulation techniques in this field. 
Section~\ref{sec:exp} delves into the specifics of our experimental setups before presenting the empirical result on benchmark test collections in Section~\ref{sec:comp-ret}.
We conclude the paper in Section~\ref{sec:conclusion} highlighting the findings considering the issues with reproducing results, and suggesting some directions to pursue further study.
\noindent
{The codes and the simulated queries are available from the following repository:} \url{https://mallpriyanshu.github.io/retrievability-survey/ecir2024}.

\section{Background and Motivation}\label{sec:background}

\subsection{Retrievability: A Measure of Accessibility}\label{subsec:retrievability}

Retrievability is a measure that evaluates the ease with which a document can be retrieved in a specific configuration of an IR system. 
Azzopardi and Vinay~\cite{azzopardi2008retrievability} introduced the concept and formally defined the retrievability score for a document $d$ in a collection $D$ as $r(d)$ with respect to a given IR system. 
Typically, the calculation of retrievability involves the execution of the following sequence of steps~\cite{thesisColin}:
$i)$ query set generation, $ii)$ system configuration and retrieval, $iii)$ computation of retrievability scores, and $iv)$ summarising the scores.

For an ideal calculation of retrievability scores, a comprehensive universe of all possible queries is needed, although obtaining this is practically unattainable.
In practice, researchers commonly select from two alternatives: either using a sufficiently large query log containing diverse user queries~\cite{roy2022studying,traub2016querylog} or simulating a set of queries from the collection that adequately represents its overall topic distribution~\cite{azzopardi2008retrievability,bache2010improving,bashir2009analyzing,bashir2009improving}.
Once an appropriate set of queries has been chosen, the second step is selecting a retrieval model with an optimal parameter setting.
Prior research has extensively examined the impact of altering system configurations~\cite{azzopardi2008retrievability,bashir2009identification,wilkie2014efficiently} where the authors conclude that changing the retrieval model can significantly impact the retrievability scores of a document collection.
The retrievals are then performed with the set of queries, and the retrievability of a document $d$ is computed in the third step as \emph{the summation of the number of times $d$ is retrieved within the top $c$ position}.
Mathematically, the retrievability $r(d)$ of a document $d$ ($d \in D)$ with respect to an IR system is computed using the formula in Equation~\ref{eq:ret}.
\begin{equation} \label{eq:ret}
    r(d) = \sum_{q \in 
    \mathsf{Q}} o_q \cdot f(k_{dq}, c)
\end{equation}
Here, $o_q$ indicates the likelihood of the query (usually considered uniform across query set) and $k_{dq}$ is the rank of $d$ for query $q$.
Calculating $r(d)$ relies on the choice of a utility function $f(.)$, either cumulative or gravity-based. Previous studies have shown a high correlation between these two measures~\cite{wilkie2013initial,wilkie2013relating}, and consequent research predominantly reports retrievability analysis using the cumulative-based function~\cite{bashir2014producing,bashir2017retrieval,traub2016querylog,wilkie_2014}.
Defined as a binary function, the cumulative utility function $f(k_{dq},c)$ is equal to $1$ if document $d$ is retrieved within the top $c$ documents for query $q$; otherwise, it is considered $0$. 
This utility function provides a straightforward interpretation of the retrievability score for each document. It simply corresponds to the number of times the document is retrieved within the top $c$ ranks. Documents beyond the top $c$ positions are disregarded, simulating a user who only examines the first $c$ results. 
Therefore, the more queries that retrieve a document within rank $c$, the higher its retrievability score.

The final step involves summarizing the retrievability scores across the collection to assess existing bias.
This quantification is commonly achieved using the Gini coefficient, a measure frequently employed in economics and social sciences to assess inequality within a population~\cite{gini1936measure}. 
Calculating the retrievability bias within a document collection, the Gini coefficient is computed as follows:
\begin{equation}\label{eq:gini}
    G = \frac{\sum_{i=1}^{N} (2i-N-1)\cdot r(d_i)}{N \sum_{j=1}^{N} r(d_j)}
\end{equation}
where $N$ represents total number of documents in the collection ($|D|$).
It provides a decimal value in the range $[0, 1]$.
A Gini coefficient of zero denotes perfect equality, indicating that all documents in the collection have an equal retrievability score according to $r(d)$. 
Conversely, a Gini coefficient of one indicates total inequality, with only one document having $r(d) = |Q|$ while all other documents have $r(d) = 0$. 
Other inequality metrics, such as Atkinson index~\cite{atkinson}, Theli index~\cite{theli}, Palma index~\cite{palma} etc., are also explored in~\cite{wilkie2015retrievability} where the authors argue that various inequality metrics generally agree on which system and settings minimize the retrievability bias and the least inequality indicated by Gini coefficient is consistent with the other metrics discussed. 
Hence, the Gini coefficient $G$ is commonly employed in subsequent works~\cite{azzopardi2008retrievability,bashir2009improving,bashir2010improving} to quantify the amount of bias in the distribution of the retrievability scores in a collection of documents.

\begin{figure}
    \centering
    \includegraphics[scale=0.47]{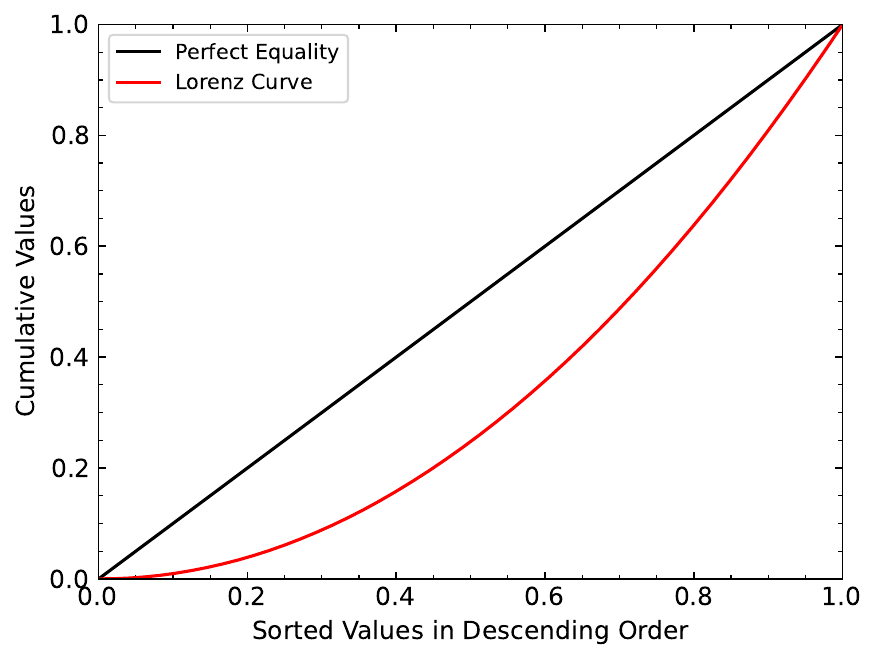}
    \caption{A sample Lorenz Curve. The values belonging to the distribution in which the disparity is to be observed are arranged in a sorted manner along the X-axis.}
    \label{fig:sample_lorenz}
\end{figure}

To visually assess the retrievability distribution in a document collection, one can calculate retrievability scores for each document using Equation~\ref{eq:ret} and plot the sorted values to create a \emph{Lorenz curve}.
This curve represents the cumulative score distribution of documents sorted by retrievability scores in ascending order.
If the retrievability scores are evenly distributed, the Lorenz Curve will be linear; a skewed curve in contrast indicates a greater level of inequality or bias. 
The Lorenz curve is a concept commonly used in economics to depict wealth distribution, where a straight line signifies equal wealth distribution across a population, while a skewed curve represents inequality, as shown in Figure~\ref{fig:sample_lorenz}.
%

Retrievability, and the underlying theory of retrievability, have found applications in various domains. For instance, it has been used in the development of inverted indexes to enhance the efficiency and performance of retrieval systems by capitalizing on terms that contribute to a document's retrievability~\cite{pickens2010reverted}. Additionally, retrievability has been leveraged to investigate bias in search engines and retrieval systems on the web~\cite{azzopardi2009search} and within patent collections~\cite{bashir2010improving}, leading to improvements in system efficiency during pruning processes~\cite{zheng2009document}.

The initial and most critical step while calculating the retrievability of a collection of documents involves query generation, followed by retrieval using the generated queries, and finally, the computation of retrievability scores based on the appearance of documents within a rank cutoff.
Prior research has extensively examined the impact of altering system configurations~\cite{azzopardi2008retrievability,bache2010improving,bashir2009identification,traub2016querylog,wilkie2014efficiently}, modifying the nature of the function~\cite{wilkie2013relating}, and varying score summarization techniques~\cite{wilkie2015retrievability}. 
However, there is limited empirical evidence concerning the influence of modifying query sets on the resultant retrievability scores.
Filling the void, in this paper, we study the effect of various query-generation techniques on the retrievability scores of documents in a collection.
We discuss some of the well-known techniques popular among researchers for query generation in the next section.

\vspace{-2mm}
\subsection{Query Generation}


The initial step in a retrievability computation typically involves obtaining a set of all possible queries addressable by the collection ($Q$ in Equation~\ref{eq:ret}).
As per Equation~\ref{eq:ret}, the most ideal calculation of $r(d)$ would require an extensive set $Q$.
In general, a query log would be the preferred option for $Q$; however, obtaining such a log can be difficult and often not feasible.
Hence, researchers have inclined towards artificial simulation of the set $Q$ from the collection.
Following the inception of the concept of retrievability in~\cite{azzopardi2008retrievability}, a popular choice among researchers in simulating $Q$ based on term statistics in the collection itself~\cite{bashir2009analyzing,bashir2009identification,bashir2009improving,bashir2010improving2,bashir2010improving,bashir2011relationship,bashir2017retrieval,noor2015evaluating}.
The approach by Azzopardi and Vinay~\cite{azzopardi2008retrievability} in generating the query set was influenced by Callan and Connel~\cite{callan2001query}. 
The authors generated a set of queries, which included single-term queries, formed by selecting each term in the vocabulary that appeared at least 5 times, and bi-term queries, formed by selecting each bigram in the collection that appeared at least 20 times. 
The list of bigrams was truncated at 20 million, and each query was issued to the system to estimate its retrievability.
Azzopardi and Bache later followed a similar methodology to generate query sets for the AP and WSJ collections by ranking the most frequent 100,000 bigrams in each collection~\cite{azzopardi2010relationship}.
The same approach was also adapted in~\cite{bache2010improving} where the authors applied retrievability to quantify the accessibility of patent documents. Although the queries used in patent search are often longer than two words, the authors choose two-word queries because this means there will be ``just one Boolean operator and it thus affords a comparison between the use of AND and OR within the various IR models''. 
In the other studies on retrievability, the prevalent choice among researchers has been the adoption of a similar approach for query generation~\cite{wilkie2013initial,wilkie2013relating,wilkie_2014,wilkie2014efficiently,wilkie2014retrievability,wilkie2015queryLength,wilkie2015retrievability}.

In their work on patent retrieval~\cite{bashir2009analyzing}, the authors created queries using terms from the \emph{claim section} of the patent documents. Single-term queries are formed from each document with a term frequency lower bound of 2.
To make longer queries, two and three-term queries are generated by the combination of the frequent terms in that document.
In their subsequent works~\cite{bashir2009identification,bashir2009improving}, the authors employed a controlled query generation technique proposed in~\cite{jordan2006using} for patent retrieval.
Subsequent works from the same research group introduced a different approach that involves generating query sets for each document in the corpus. They select terms that appear more than once within the document, resulting in four query subsets: single terms, combinations of two terms, combinations of three terms, and combinations of four terms~\cite{bashir2010improving2,bashir2010improving,bashir2011relationship}. 
Some further truncation of query set was adapted in~\cite{bashir2012improvingRetrievabilityCluster,bashir2014estimating,bashir2014producing,bashir2014automatic,bashir2017retrievalModel} where queries were generated with combinations of terms that appear more than once in the document. Stopwords and terms having document frequency $>$ 25\% of collection size were removed.

Assessing document retrievability in a collection necessitates a comprehensive set of queries that cover all aspects of the collection. Ideally, this would involve an extensive query log spanning a significant time period. However, acquiring such a log is often costly and unavailable, prompting researchers to resort to simulated queries.
Despite the challenges, some research has been conducted using real user query logs acquired from certain closed domains.
The authors in~\cite{traub2018impact,traub2016querylog} use a set of real queries submitted to the National Library of the Netherlands, between March 2015 to July 2015. Additionally, to make the results comparable to~\cite{azzopardi2008retrievability}, the authors create a set of simulated queries: they select the top 2 million unigrams and bigrams respectively as single term and two term queries. All the queries were finalized after stemming and removal of stopwords and operators. 
On comparing the characteristics of simulated queries to those formed by real user queries, the authors report significant differences regarding the composition, the number of unique terms used, and the presence of named entities: real user queries exhibited a considerably higher fraction of named entities compared to the simulated query set.
Additionally, due to the considerably low number of real queries in contrast to the vast size of the collection (exceeding 102 million), the authors acknowledged the presence of a varying document-query ratio compared to those reported in~\cite{azzopardi2008retrievability}.
Authors in~\cite{amin2023} employ a transformer-based generative model docT5query~\cite{doct5query} for generating queries and conclude that BERT-based rankers demonstrate less retrievability bias than user-given queries.
Recently, Roy et al.~\cite{roy2022studying} studied the retrievability distribution in an integrated retrieval system containing documents in multiple categories~\cite{roy2022studying,roy2023retrievability}.
However, this work does not conduct a direct comparison of the retrievability scores (and their distributions) when computed using real and simulated query sets.

The contrasting nature of simulated queries, when compared to real queries, is briefly reported in~\cite{traub2016querylog}. 
The various techniques used for generating simulated queries illustrate the diversity in approaches taken at the initial stage of a retrievability analysis by different groups of researchers.
This also highlights the absence of a well-defined structure for such an analysis paving a potential issue with reproducing results.
It prompts the question: can we anticipate a similar distribution of retrievability scores when the query set is altered?
Answering that question, in this paper, we present a systematic comparison of the retrievability scores computed based on several of these techniques together with real-user query logs. 
According to the observation in~\cite{traub2016querylog}, the real queries contain a significantly higher fraction of named entities compared to the simulated query set.
Inspired by this observation, we additionally introduce a rule-based query simulation technique for retrievability assessment, further contributing to the examination of this diversity in approaches.
To maintain the primary narrative of the paper, we have detailed the method in the Appendix~\ref{appen:rule_simulation}.

\section{Experimental Setup}\label{sec:exp}

As presented in Section~\ref{sec:background}, the set of queries employed in computing the retrievability plays an important role in the computation and our focus of this paper is to foresee whether alteration in this set can affect the final score,
potentially leading to challenges in reproducing results for different collections and retrieval models.
In this section, we present the datasets and the query sets that are used in this paper to verify this empirically with real data.

\subsection{Datasets}\label{subsec:dataset}

We perform the experiments of the retrievability score distributions on two benchmark retrieval datasets as well as on a diverse encyclopedia of documents.
Particularly, the TREC ad-hoc Robust collection~\cite{overviewRobust} containing news articles, WT10g collection~\cite{wt10g} with web pages, together with the English Wikipedia article dump of February, 2023 are employed in this study\footnote{\url{https://dumps.wikimedia.org/enwiki/}}.
All the datasets are preprocessed with stopword removal following SMART stopword list~\footnote{\url{https://tinyurl.com/smart-stopword}} and stemmed using the Porter Stemmer before indexing them using Apache Lucene.
Overall statistics of the datasets are presented in Table~\ref{tab:dataset}.

\begin{table}[t]
    \centering
    \caption{Basic statistics of datasets used for empirical verification of the variation in the computed retrievability scores when different query sets are employed.}
    \label{tab:dataset}
    \begin{tabular}{lrcr} \toprule
        \textbf{Dataset} & \textbf{~~~\# documents} & \textbf{~~Collection Type} & \textbf{~~~\# terms} \\ \hline
        \textbf{TREC Robust} & 528,155 & ~News articles~ & 1,502,031 \\
        \textbf{WT10g} & 1,692,096 & ~Web pages~ & 9,674,707 \\
        \textbf{Wikipedia} & 6,584,626 & Encyclopedia & 18,797,260 \\ \hline
    \end{tabular}
    \vspace{-4mm}
\end{table}

\subsection{Simulated Queries}\label{subsec:simu-queries}
In order to foresee the variation, if any, in the retrievability distribution of documents in a collection, we use the following artificial query generation techniques:
\begin{enumerate}
    \item \textbf{Simulated Query set 1 ($\mathsf{SQ_{1}}$):} Following the suggestion in~\cite{azzopardi2008retrievability}, the first set of the queries with which the retrievals are performed is formed by a sampling method proposed in~\cite{callan2001query}.
In this process, the terms undergo analysis and filtering, including stemming using the Porter stemmer and removing stopwords. Terms occurring more than 5 times in the collection are treated as single-term queries. Additionally, two-term queries are formed by pairing consecutively occurring terms that have a collection frequency of at least 20 times. These resulting bigrams are then ranked by occurrence, and the top two million are selected as the final set of two-term queries.

    \item \textbf{Simulated Query set 2 ($\mathsf{SQ_{2}}$):} The second technique is based on~\cite{bashir2014producing}. Specifically, queries are generated with the combinations of terms that appear more than once in the document. Stopwords and other terms with document frequency $>$ 25\% of collection size are removed. Queries up to 4 terms are formed using the boolean AND operator, with duplicate queries being removed.

    \item \textbf{Simulated Query set 3 ($\mathsf{SQ_{3}}$):} In this approach (based on~\cite{jordan2006using}), a set of related documents is chosen as the source document set with clustering. Language models are defined for both the source documents and the corpus, with relative entropy used for comparison. Terms in the vocabulary are sorted based on their contribution to relative entropy, serving as a term discrimination scoring function. The most discriminating term initiates the query construction process, forming single-term and two-term queries for different query sets. The process continues by identifying next most informative term and generating subsequent queries until all terms have been used.

    \item \textbf{Rule-based Simulated Query set ($\mathsf{RSQ}$):} 
    Building upon the insight in~\cite{traub2016querylog}, that real queries contain a notably greater proportion of named entities compared to simulated queries, we introduce a rule-based query simulation technique for retrievability assessment. 
    The approach is presented in Appendix~\ref{appen:rule_simulation}.
\end{enumerate}

\begin{table}[t]
    \centering
    \caption{Number of queries generated for each dataset using the individual query generation techniques.
    Note that, the simulated queries are generated from each dataset and applied in that particular dataset for the retrievability computation.
    AOL query set is filtered separately for each dataset as discussed in Section~\ref{subsec:real-queries}.
    }
    \label{tab:query}
    \begin{tabular}{l|c|r|r|r}
    \toprule
    & \multirow{2}{*}{\textbf{Query Type}} & \multicolumn{3}{c}{\textbf{Number of queries}} \\ \cmidrule{3-5}
    & & {\textbf{~Robust04~}} & {\textbf{~~WT10g~}} & {\textbf{~~Wikipedia~~}} \\
    \hline
    \textbf{$\mathsf{SQ_{1}}$}~\cite{azzopardi2008retrievability} & Simulated &  1061184~ & 3902575~ & 5320001~~ \\
    \textbf{$\mathsf{SQ_{2}}$}~\cite{bashir2014producing} & Simulated & 6903196~ & 18724509~ & 84920098~~ \\
    \textbf{$\mathsf{SQ_{3}}$}~\cite{jordan2006using} & Simulated & 1644533~ & 7346091~  & 26272419~~ \\
    \textbf{$\mathsf{RSQ}$} & Simulated & 3900000~ & 4500000~ & 4550000~~ \\
    \textbf{$\mathsf{AOL}$}~\cite{aol-log} & Real Query & 3885206~ & 4302650~ & 4550977~~ \\
    \bottomrule
    \end{tabular}
\end{table}

\subsection{Real Queries}\label{subsec:real-queries}
A majority of the works done in the domain of retrievability employ queries generated using simulation techniques. 
In this work, we use a general web search query log to make our study conclusive.
Specifically, we utilize the AOL query log~\cite{aol-log} that contains actual user query click data spanning from March 1st, 2006, to May 31st, 2006. 
It sparked significant controversy within the media, primarily due to privacy-related concerns~\cite{aol-news1,aol-news2,aol-news3}.
Although the query log is more than a decade old, it is still widely used in recent literature~\cite{dehghani2017aol,ahmad2019aol,ma2020aol,kang2021aol,ahu2021aol,macavaney_use_aol}.

To prepare the real query set for our experiments, the initial step involves extracting all unique queries from the AOL query log.
Queries containing periods are eliminated to filter out website links. 
Given the distinct temporal scopes of our datasets and query sets, a conspicuous challenge arises in the form of topic and vocabulary mismatch. 
To mitigate this challenge as much as possible, we adopt an exclusionary approach:
a query is selected if all of its terms are present in the corpus vocabulary.
This strategic refinement serves to align the queries closely with the content of the target collection. 
Note that, the AOL queries are refined for each collection separately following the above exclusionary approach.
Table~\ref{tab:query} contains the individual count of queries finalized for each dataset.
%

\subsection{Retrieval Model}
In accordance with the insights and recommendations put forth in~\cite{azzopardi2008retrievability}, we only use BM25 as the model for  performing the retrieval. 
Specifically, we use the implementation available in Apache Lucene.
To fine-tune the BM25 parameters, we utilize the TREC ad-hoc retrieval topic set~\cite{overviewRobust}, which is then applied on the Robust dataset for the retrievability study. 
Similarly, the TREC web topic sets~\cite{wt10g} are employed for parameter tuning, and these adjusted parameters are subsequently applied to the WT10g and Wikipedia datasets.

\begin{table}[t]
    \centering
    \caption{Gini coefficient of Retrievability score distribution when computed using different query generation techniques. The highest and the lowest Gini values for each dataset are marked respectively with a superscript $^\dagger$ and $^*$. }
    {
    \begin{tabular}{l|ccccc}
    \toprule
    \textbf{Dataset} & {\textbf{$\mathsf{SQ_{1}}$}} & {\textbf{$\mathsf{SQ_{2}}$}} & {\textbf{$\mathsf{SQ_{3}}$}} & {\textbf{$\mathsf{RSQ}$}} & {$\mathsf{AOL}$}\\ \hline
    \textbf{Robust} & $~~0.3307~~$ & $~~0.4556~~$ & $~~0.4300~~$ & $~~0.3052^*~~$ & $~~{0.6032}^\dagger~~$ \\
    \textbf{WT10g}       & 0.5371 & 0.6391 & 0.6359 & $0.5009^*$ & $0.6541^\dagger$\\
    \textbf{Wikipedia~~} & 0.5380 & 0.6210 & 0.6290 & $0.4820^*$ & $0.6798^\dagger$ \\
    \hline
    \end{tabular}
    }
    \label{tab:gini}
\end{table}

\begin{figure}[t]
    \centering
    \includegraphics[scale=0.45]{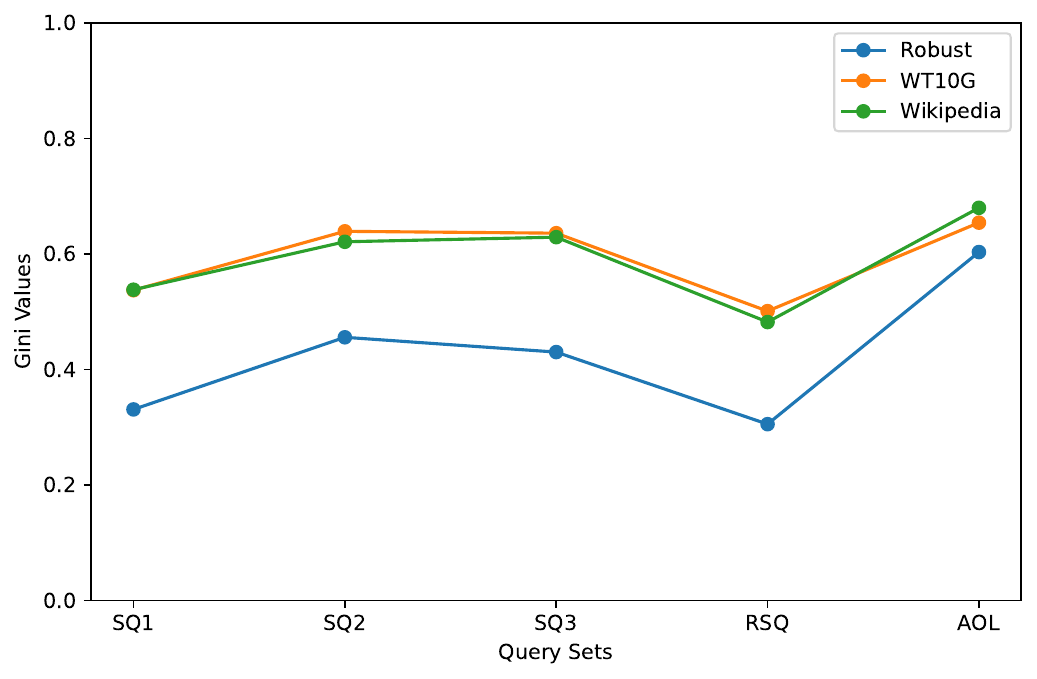}
    \caption{Gini coefficient for each dataset when the retrievability scores are calculated using different query sets as discussed in Section~\ref{sec:exp}.}
    \label{fig:all-gini}
\end{figure}

\section{Comparing Retrievability}\label{sec:comp-ret}
\subsection{Gini Coefficient and Lorenz Curve}\label{subsec:comp-gini}

\begin{figure*}[t]
  \centering
  \begin{subfigure}{0.5115\textwidth}
    \includegraphics[width=\linewidth]{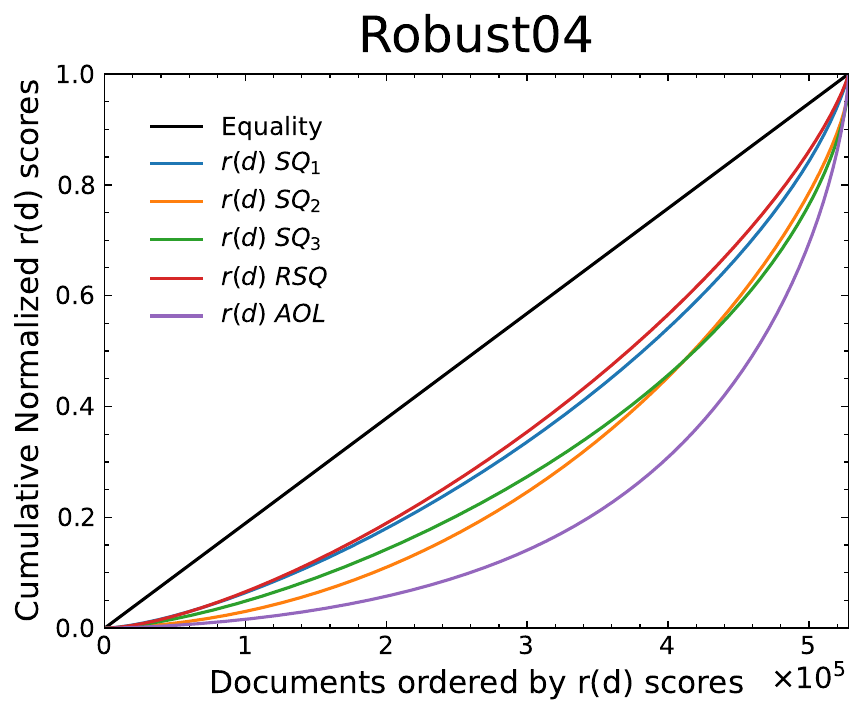}
  \end{subfigure}
  \hspace{-2.5mm}
  \begin{subfigure}{0.4875\textwidth}
    \includegraphics[width=\linewidth]{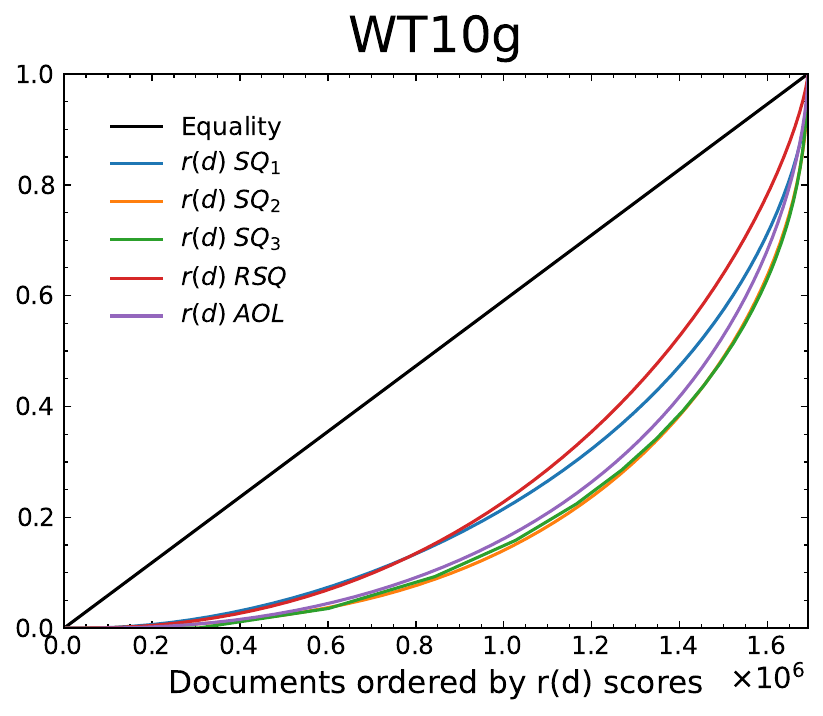}
  \end{subfigure}
  \hspace{-2.5mm}
  \begin{subfigure}{0.48\textwidth}
    \includegraphics[width=\linewidth]{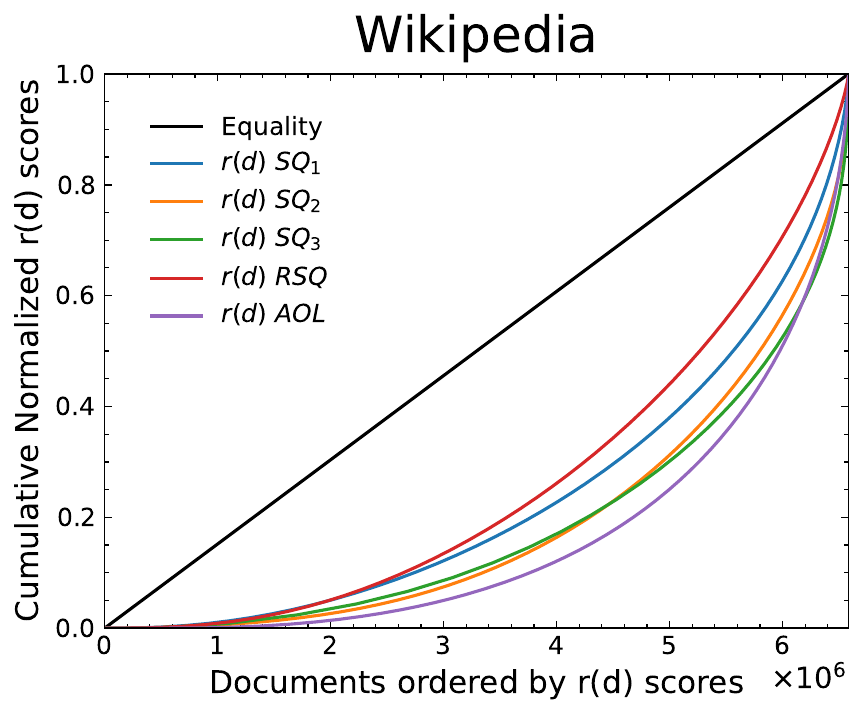}
  \end{subfigure}
  \caption{Lorenz curve with retrievability computed with various query sets on different collections.} \label{fig:lorenz}
\end{figure*}

We present the Gini coefficient
for each dataset utilizing all five query sets in Table~\ref{tab:gini}.
The variation in Gini values is graphically presented in Figure~\ref{fig:all-gini}.
In the table, the least disparity is observed across all the datasets when utilizing the rule-based simulation for query generation (denoted as $\mathsf{RSQ}$ in Table~\ref{tab:gini}). 
The distributions of retrievability values are presented using Lorenz curve in Figure~\ref{fig:lorenz}.
Furthermore, the original query simulation technique proposed by~\cite{azzopardi2008retrievability} (indicated by $\mathsf{SQ_{1}}$ in Table~\ref{tab:gini}) follows a similar trend.
The other two simulation techniques ($\mathsf{SQ_{2}}$ and $\mathsf{SQ_{3}}$) exhibit higher disparity.
In contrast, a significant disparity in retrievability scores across all the datasets is observed when the real-query log ($\mathsf{AOL}$) is used.
These results lead to the following conclusions:
%
\begin{itemize}
    \item The web collection WT10g and the encyclopedia Wikipedia exhibit much more diverse retrievability scores as compared to the relatively smaller news collection TREC Robust across all the query sets.
    \item The approach in~\cite{azzopardi2008retrievability} selects terms from the collection based on their frequency, including non-informative queries ($\mathsf{SQ_1}$). This method results in a more evenly distributed query generation covering a wide range of documents in the collection. As a result, it tends to produce lower Gini values compared to other query sets.
    \item Realistic queries generated using the rule-based approach ($\mathsf{RSQ}$) result in even lower Gini coefficients across all datasets.
    \item AOL query logs mostly consist of known item searches rather than exploratory ones leading to a significant disparity in computed retrievability scores, with Gini coefficients consistently exceeding $0.6$ across all datasets.
\end{itemize}

\begin{table*}[t]
    \centering
    \caption{Correlation between retrievability scores when 
      computed using the real query set (AOL) and 
      the four simulated query sets (discussed in
      Section~\ref{subsec:simu-queries}).
      Upper triangle contains Pearson's $r$ while lower
      triangle reports Kendall's rank correlation $\tau$.
      All differences statistically significant with $p
      < 0.05$.
      Highest (in bold) and lowest (superscripts $^*$) correlations for each dataset are highlighted.
    }
    \label{tab:correl}
    \resizebox{0.83\columnwidth}{!}
    {
\begin{tabular}{||c||ccccc||}
  \hline
  & \multicolumn{5}{c||}{\textbf{Robust04}} \\ \hline 
  & 
  \multicolumn{1}{c}{\textbf{~~~~SQ${_{1}~~~~}$}} & \multicolumn{1}{c}{\textbf{~~~~SQ${_{2}~~~~}$}} & \multicolumn{1}{c}{\textbf{~~~~SQ${_{3}~~~~}$}} & \multicolumn{1}{c}{\textbf{~~~~RSQ${~~~~}$}} & \multicolumn{1}{c||}{\textbf{~~~~AOL${~~~~}$}}  \\\hline
  
\textbf{SQ${_{1}}$}     & \multicolumn{1}{c}{\cellcolor{gray!25}}       & \multicolumn{1}{c}{0.0923}  & \multicolumn{1}{c}{-0.0025} & \multicolumn{1}{c}{~\textbf{0.7549}} & 0.1250                   \\ \hline
\textbf{SQ${_{2}}$} & \multicolumn{1}{c}{0.1758} & \multicolumn{1}{c}{\cellcolor{gray!25}}        & \multicolumn{1}{c}{-0.0004$^*$} & \multicolumn{1}{c}{0.1461} & 0.0033                \\ \hline
\textbf{SQ${_{3}}$} & \multicolumn{1}{c}{0.0108} & \multicolumn{1}{c}{-0.0040$^*$} & \multicolumn{1}{c}{\cellcolor{gray!25}}        & \multicolumn{1}{c}{0.0030} & 0.0165 \\ \hline
\textbf{RSQ} & \multicolumn{1}{c}{~\textbf{0.5617}} & \multicolumn{1}{c}{0.3023}  & \multicolumn{1}{c}{0.0233}  & \multicolumn{1}{c}{\cellcolor{gray!25}}       & 0.2076   \\ \hline
\textbf{~~~AOL~~~} & \multicolumn{1}{c}{0.1413} & \multicolumn{1}{c}{-0.0174} & \multicolumn{1}{c}{0.0321}  & \multicolumn{1}{c}{0.1811} & \multicolumn{1}{l||}{\cellcolor{gray!25}}  \\ \hline \hline

  & \multicolumn{5}{c||}{\textbf{WT10g}} \\ \hline 

\textbf{SQ${_{1}}$}  & \multicolumn{1}{c}{\cellcolor{gray!25}}       & \multicolumn{1}{c}{0.1206} & \multicolumn{1}{c}{0.2094} & \multicolumn{1}{c}{0.3103} & 0.1100                \\ \hline
\textbf{SQ${_{2}}$}                   & \multicolumn{1}{c}{0.3575} & \multicolumn{1}{c}{\cellcolor{gray!25}}       & \multicolumn{1}{c}{~\textbf{0.5503}} & \multicolumn{1}{c}{0.1931} & \multicolumn{1}{c||}{~0.0900$^*$} \\ \hline
\textbf{SQ${_{3}}$} & \multicolumn{1}{c}{0.3223} & \multicolumn{1}{c}{0.3047} & \multicolumn{1}{c}{\cellcolor{gray!25}}       & \multicolumn{1}{c}{0.2535} & 0.1795                    \\ \hline
\textbf{RSQ}       & \multicolumn{1}{c}{~\textbf{0.5672}} & \multicolumn{1}{c}{0.5312} & \multicolumn{1}{c}{0.3189} & \multicolumn{1}{c}{\cellcolor{gray!25}}       & 0.3135 \\ \hline
\textbf{AOL} & \multicolumn{1}{c}{0.3731} & \multicolumn{1}{c}{~0.2977$^*$} & \multicolumn{1}{c}{0.3830} & \multicolumn{1}{c}{0.4126} & \multicolumn{1}{l||}{\cellcolor{gray!25}}    \\ \hline \hline

  & \multicolumn{5}{c||}{\textbf{Wikipedia}} \\ \hline 

\textbf{SQ${_{1}}$} & \multicolumn{1}{c}{\cellcolor{gray!25}}       & \multicolumn{1}{c}{0.1586} & \multicolumn{1}{c}{0.2613} & \multicolumn{1}{c}{0.4768} & 0.2379                   \\ \hline
\textbf{SQ${_{2}}$} & \multicolumn{1}{c}{0.3508} & \multicolumn{1}{c}{\cellcolor{gray!25}}       & \multicolumn{1}{c}{~\textbf{0.7019}} & \multicolumn{1}{c}{0.1599} & 0.0598                   \\ \hline
\textbf{SQ${_{3}}$} & \multicolumn{1}{c}{0.3897} & \multicolumn{1}{c}{0.5116} & \multicolumn{1}{c}{\cellcolor{gray!25}}       & \multicolumn{1}{c}{0.1048} & \multicolumn{1}{c||}{~0.0308$^*$}                   \\ \hline
\textbf{RSQ}                & \multicolumn{1}{c}{~\textbf{0.5350}} & \multicolumn{1}{c}{0.4867} & \multicolumn{1}{c}{0.3739} & \multicolumn{1}{c}{\cellcolor{gray!25}}       & 0.3065                   \\ \hline
\textbf{AOL} & \multicolumn{1}{c}{0.3603} & \multicolumn{1}{c}{0.3074} & \multicolumn{1}{c}{~0.2338$^*$} & \multicolumn{1}{c}{0.4194} & \multicolumn{1}{l||}{\cellcolor{gray!20}}    \\ \hline 
\end{tabular}
}
\end{table*}



\subsection{Comparing Correlation}
Table~\ref{tab:correl} reports the degree of variation in retrievability scores when comparing the simulation-based query generation techniques ($\mathsf{SQ_1}$, $\mathsf{SQ_2}$, $\mathsf{SQ_3}$, and $\mathsf{RSQ}$) along with the AOL query log.
We compute the Pearson's $r$ between the distribution of each pair of retrievability scores computed using the different query sets and report in the upper half of Table~\ref{tab:correl}.
Kendall's $\tau$ is presented in the lower half of the table which is computed using the document rank arranged following their retrievability scores.
%
The table shows negligible correlations between scores computed using almost any query set pairs.
Interestingly, the highest correlations, in terms of both Pearson's $r$ and Kendall's $\tau$ occur when using the rule-based technique $\mathsf{RSQ}$ (Appendix~\ref{appen:rule_simulation}). 
Further, the massive difference in the highest and lowest correlations validates that distinct query sets lead to divergent retrievability scores, affirming both \textbf{RQ1} and \textbf{RQ2}.

\section{Conclusion}\label{sec:conclusion}

The examination of retrievability has predominantly relied on the use of simulated queries since its inception with few works using real query logs. 
However,
a comprehensive comparative analysis of the computed values has been absent in the existing literature. 
Bridging this gap, in this paper, we examine the impact of five distinct and diverse query sets, ranging from simulated queries to actual user query logs, on the computed retrievability scores when the retrieval is performed using BM25.

Our findings on diverse datasets, including news, web, and Wikipedia collections expose substantial variations in computed retrievability scores across query sets, providing an affirmative response to \textbf{RQ1} and underscoring the reproducibility challenges associated with retrievability scores.
Among the popular query simulation techniques for retrievability studies, the query set generated following the method in~\cite{azzopardi2008retrievability} ($\mathsf{SQ_1}$) exhibits the least disparity in computed retrievability scores.
In contrast, the AOL query log generates the highest disparity.
Furthermore, the utilization of a rule-based simulation technique ($\mathsf{RSQ}$) results in the least inequality in the retrievability scores among all. These observations reveal a high sensitivity of these scores with respect to the query generation methods employed. 
Our research demonstrates that the selection of a query set directly impacts computed retrievability scores, regardless of the nature of the collection. This provides a conclusive response to \textbf{RQ2}.
These findings emphasize the need for standardization of query set construction in retrievability studies and contribute to a deeper understanding of the nuances of accessibility within the field of information retrieval.
In previous research on retrievability, various studies utilized multiple retrieval models for a single query set per collection. 
As part of our future work, we will explore the impact of diverse retrieval models on the retrievability scores with different query sets.

\appendix
\clearpage
\section{Appendix - A POS-based Query Generation Technique}\label{appen:rule_simulation}


Identifying collocations in a text corpus typically involves counting co-occurring word pairs, revealing words that go beyond their individual meanings. Relying solely on the most frequent bigrams often yields uninteresting results, as many of them consist of function words, offering limited insights. To improve collocation quality, Justeson and Katz~\cite{justeson1991co} introduced a simple yet effective heuristic. They apply a part-of-speech filter to candidate phrases, preserving patterns likely to represent genuine `phrases' rather than random word combinations. This approach enhances the meaningfulness of the collocation identification process.


We apply this approach for query generation improving the relevance and effectiveness of the generated queries.

\begin{enumerate}
    \item \textbf{Perform Part-of-Speech (POS) tagging}: Initially, we employ POS tagging on all the documents within the collection. This step assigns appropriate grammatical tags to each word, facilitating the subsequent identification of N-grams.
    \item \textbf{Extract N-grams}: N-grams, where N represents the desired length of the word sequences, are extracted from the POS-tagged documents. In our case, we consider N to range from 1 to 4, enabling the identification of unigrams, bigrams, trigrams, and quadgrams.
    \item \textbf{Select N-grams with `query-like' POS tag patterns}: From the pool of extracted N-grams, we apply Justeson and Katz's~\cite{justeson1991co} recommended POS tag patterns to filter and retain N-grams that exhibit patterns resembling queries. The specific POS tag patterns for each N-gram type are provided in Table \ref{tab:pos-tag}. Tag patterns for Quadgrams are proposed by us heuristically from our observations.
\end{enumerate}
\vspace{-5mm}
\begin{table}[h]
\centering
\tiny
\caption{POS Tag rules for N-gram query generation}
\label{tab:pos-tag}
\resizebox{0.99\columnwidth}{!}{
\begin{tabular}{|l|l|l|l|}
\hline
\textbf{unigram}  & \textbf{bigram}  & \textbf{trigram}       & \textbf{quadgram}     \\ \hline
noun            & adj noun & adj adj noun & noun verb adp noun \\ 
                & noun noun & adj noun noun & adj noun adj noun \\ 
                & & noun adj noun & noun adp adj noun \\ 
                & & noun noun noun & noun noun adp noun \\ 
                & & noun adp noun & noun verb noun noun \\ 
                & & & adv adj noun noun \\ 
                & & & adj noun verb noun \\ 
                & & & noun adj noun noun \\ \hline
\end{tabular}
}
\end{table}

Subsequently, the resulting list of N-grams is sorted in descending order based on their occurrence frequencies. To ensure a manageable and relevant set of queries, we truncate the list at specific thresholds. These thresholds are determined by drawing inspiration from the frequency distribution of queries found in the AOL query set~\cite{aol-log}. We aim to maintain a proportional ratio between the selected N-grams and the query frequencies observed in the AOL real query set, preserving a close alignment with real-world query usage patterns.

%
%
%
\bibliographystyle{splncs04}
\bibliography{samplepaper}

\begin{thebibliography}{10}
\providecommand{\url}[1]{\texttt{#1}}
\providecommand{\urlprefix}{URL }
\providecommand{\doi}[1]{https://doi.org/#1}

\bibitem{ahmad2019aol}
Ahmad, W.U., Chang, K.W., Wang, H.: Context attentive document ranking and
  query suggestion. In: Proceedings of the 42nd International ACM SIGIR
  Conference on Research and Development in Information Retrieval. p.
  385–394. SIGIR'19, Association for Computing Machinery, New York, NY, USA
  (2019). \doi{10.1145/3331184.3331246},
  \url{https://doi.org/10.1145/3331184.3331246}

\bibitem{amin2023}
Amin~Abolghasemi, Suzan~Verberne, A.A., Azzopardi, L.: Retrievability bias
  estimation using synthetically generated queries. In: Proceedings of the
  First Workshop on Generative Information Retrieval - GenIR@SIGIR2023 held in
  conjunction with SIGIR 2023. GenIR@SIGIR2023 (2023),
  \url{https://coda.io/@sigir/gen-ir/accepted-papers-17}

\bibitem{aol-news3}
Anderson, N.: The ethics of using aol search data
  \url{https://arstechnica.com/uncategorized/2006/08/7578/}

\bibitem{atkinson}
Atkinson, A.B.: On the measurement of inequality. Journal of Economic Theory
  \textbf{2}(3),  244--263 (1970).
  \doi{https://doi.org/10.1016/0022-0531(70)90039-6},
  \url{https://www.sciencedirect.com/science/article/pii/0022053170900396}

\bibitem{azzopardi2010relationship}
Azzopardi, L., Bache, R.: On the relationship between effectiveness and
  accessibility. In: Proceedings of the 33rd international ACM SIGIR conference
  on Research and development in information retrieval. pp. 889--890 (2010)

\bibitem{azzopardi2009search}
Azzopardi, L., Owens, C.: Search engine predilection towards news media
  providers. In: Proceedings of the 32nd international ACM SIGIR conference on
  Research and development in information retrieval. pp. 774--775 (2009)

\bibitem{azzopardi2008accessibility}
Azzopardi, L., Vinay, V.: Accessibility in information retrieval. In: Advances
  in Information Retrieval: 30th European Conference on IR Research, ECIR 2008,
  Glasgow, UK, March 30-April 3, 2008. Proceedings 30. pp. 482--489. Springer
  (2008)

\bibitem{azzopardi2008retrievability}
Azzopardi, L., Vinay, V.: Retrievability: An evaluation measure for higher
  order information access tasks. In: Proceedings of the 17th ACM Conference on
  Information and Knowledge Management. p. 561–570. CIKM '08, Association for
  Computing Machinery, New York, NY, USA (2008). \doi{10.1145/1458082.1458157},
  \url{https://doi.org/10.1145/1458082.1458157}

\bibitem{bache2010improving}
Bache, R., Azzopardi, L.: Improving access to large patent corpora. Trans.
  Large Scale Data Knowl. Centered Syst.  \textbf{2},  103--121 (2010).
  \doi{10.1007/978-3-642-16175-9\_4},
  \url{https://doi.org/10.1007/978-3-642-16175-9\_4}

\bibitem{aol-news1}
Barbaro, Michael; Zeller~Jr, T.: A face is exposed for aol searcher no. 4417749
  \url{https://www.nytimes.com/2006/08/09/technology/09aol.html}

\bibitem{bashir2012improvingRetrievabilityCluster}
Bashir, S.: Improving retrievability with improved cluster-based
  pseudo-relevance feedback selection. Expert Syst. Appl.  \textbf{39}(8),
  7495--7502 (2012). \doi{10.1016/j.eswa.2012.01.041},
  \url{https://doi.org/10.1016/j.eswa.2012.01.041}

\bibitem{bashir2014estimating}
Bashir, S.: Estimating retrievability ranks of documents using document
  features. Neurocomputing  \textbf{123},  216--232 (2014)

\bibitem{bashir2014producing}
Bashir, S., Khattak, A.S.: Producing efficient retrievability ranks of
  documents using normalized retrievability scoring function. Journal of
  Intelligent Information Systems  \textbf{42},  457--484 (2014).
  \doi{10.1007/s10844-013-0274-3},
  \url{https://doi.org/10.1007/s10844-013-0274-3}

\bibitem{bashir2009analyzing}
Bashir, S., Rauber, A.: Analyzing document retrievability in patent retrieval
  settings. In: Database and Expert Systems Applications: 20th International
  Conference, DEXA 2009, Linz, Austria, August 31--September 4, 2009.
  Proceedings 20. pp. 753--760. Springer (2009)

\bibitem{bashir2009identification}
Bashir, S., Rauber, A.: Identification of low/high retrievable patents using
  content-based features. In: Proceedings of the 2nd international workshop on
  Patent information retrieval. pp. 9--16 (2009)

\bibitem{bashir2009improving}
Bashir, S., Rauber, A.: Improving retrievability of patents with cluster-based
  pseudo-relevance feedback documents selection. In: Proceedings of the 18th
  ACM conference on Information and knowledge management. pp. 1863--1866 (2009)

\bibitem{bashir2010improving2}
Bashir, S., Rauber, A.: Improving retrievability and recall by automatic corpus
  partitioning. Transactions on large-scale data-and knowledge-centered systems
  II pp. 122--140 (2010)

\bibitem{bashir2010improving}
Bashir, S., Rauber, A.: Improving retrievability of patents in prior-art
  search. In: Advances in Information Retrieval: 32nd European Conference on IR
  Research, ECIR 2010, Milton Keynes, UK, March 28-31, 2010. Proceedings 32.
  pp. 457--470. Springer (2010)

\bibitem{bashir2011relationship}
Bashir, S., Rauber, A.: On the relationship between query characteristics and
  ir functions retrieval bias. Journal of the American Society for Information
  Science and Technology  \textbf{62}(8),  1515--1532 (2011)

\bibitem{bashir2014automatic}
Bashir, S., Rauber, A.: Automatic ranking of retrieval models using
  retrievability measure. Knowledge and information systems  \textbf{41},
  189--221 (2014)

\bibitem{bashir2017retrieval}
Bashir, S., Rauber, A.: Retrieval models versus retrievability. Current
  Challenges in Patent Information Retrieval pp. 185--212 (2017)

\bibitem{bashir2017retrievalModel}
Bashir, S., Rauber, A.: Retrieval Models Versus Retrievability, pp. 185--212.
  Springer Berlin Heidelberg, Berlin, Heidelberg (2017).
  \doi{10.1007/978-3-662-53817-3\_7},
  \url{https://doi.org/10.1007/978-3-662-53817-3\_7}

\bibitem{ludovico2023bias}
Boratto, L., Faralli, S., Marras, M., Stilo, G. (eds.): Advances in Bias and
  Fairness in Information Retrieval. Springer Nature Switzerland (2023).
  \doi{10.1007/978-3-031-37249-0},
  \url{https://doi.org/10.1007/978-3-031-37249-0}

\bibitem{callan2001query}
Callan, J., Connell, M.: Query-based sampling of text databases. ACM
  Transactions on Information Systems (TOIS)  \textbf{19}(2),  97--130 (2001)

\bibitem{dehghani2017aol}
Dehghani, M., Zamani, H., Severyn, A., Kamps, J., Croft, W.B.: Neural ranking
  models with weak supervision. In: Proceedings of the 40th International ACM
  SIGIR Conference on Research and Development in Information Retrieval. p.
  65–74. SIGIR '17, Association for Computing Machinery, New York, NY, USA
  (2017). \doi{10.1145/3077136.3080832},
  \url{https://doi.org/10.1145/3077136.3080832}

\bibitem{ekstrand2022foundations}
Ekstrand, M.D., Das, A., Burke, R., Diaz, F.: Fairness in information access
  systems. Foundations and Trends® in Information Retrieval  \textbf{16}(1-2),
   1--177 (2022). \doi{10.1561/1500000079},
  \url{http://dx.doi.org/10.1561/1500000079}

\bibitem{gini1936measure}
Gini, C.: On the measure of concentration with special reference to income and
  statistics. Colorado College Publication, General Series  \textbf{208}(1),
  73--79 (1936)

\bibitem{aol-news2}
Hafner, K.: Tempting data, privacy concerns; researchers yearn to use aol logs,
  but they hesitate
  \url{https://www.nytimes.com/2006/08/23/technology/23search.html}

\bibitem{wt10g}
Hawking, D.: Overview of the {TREC-9} web track. In: Voorhees, E.M., Harman,
  D.K. (eds.) Proceedings of The Ninth Text REtrieval Conference, {TREC} 2000,
  Gaithersburg, Maryland, USA, November 13-16, 2000. {NIST} Special
  Publication, vol. 500-249. National Institute of Standards and Technology
  {(NIST)} (2000), \url{http://trec.nist.gov/pubs/trec9/papers/web9.pdf}

\bibitem{theli}
Johnston, J.: {H. Theil. Economics and Information Theory}. The Economic
  Journal  \textbf{79}(315),  601--602 (09 1969). \doi{10.2307/2230396},
  \url{https://doi.org/10.2307/2230396}

\bibitem{jordan2006using}
Jordan, C., Watters, C., Gao, Q.: Using controlled query generation to evaluate
  blind relevance feedback algorithms. In: Proceedings of the 6th ACM/IEEE-CS
  joint conference on Digital libraries. pp. 286--295 (2006)

\bibitem{justeson1991co}
Justeson, J.S., Katz, S.M.: Co-occurrences of antonymous adjectives and their
  contexts. Computational Linguistics  \textbf{17}(1),  1--20 (1991),
  \url{https://aclanthology.org/J91-1001}

\bibitem{kang2021aol}
Kang, Y.M., Liu, W., Zhou, Y.: Queryblazer: Efficient query autocompletion
  framework. In: Proceedings of the 14th ACM International Conference on Web
  Search and Data Mining. p. 1020–1028. WSDM '21, Association for Computing
  Machinery (2021). \doi{10.1145/3437963.3441725}

\bibitem{ma2020aol}
Ma, Z., Dou, Z., Bian, G., Wen, J.R.: Pstie: Time information enhanced
  personalized search. In: Proceedings of the 29th ACM International Conference
  on Information \& Knowledge Management. p. 1075–1084. CIKM '20, Association
  for Computing Machinery (2020). \doi{10.1145/3340531.3411877}

\bibitem{macavaney_use_aol}
MacAvaney, S., Macdonald, C., Ounis, I.: Reproducing personalised session
  search over the aol query log. In: Advances in Information Retrieval: 44th
  European Conference on IR Research, ECIR 2022, Stavanger, Norway, April
  10–14, 2022, Proceedings, Part I. p. 627–640. Springer-Verlag, Berlin,
  Heidelberg (2022). \doi{10.1007/978-3-030-99736-6\_42}

\bibitem{thesisColin}
McLellan, C.: The relationship between retrievability bias and retrieval
  performance. Ph.D. thesis, University of Glasgow, {UK} (2019),
  \url{https://ethos.bl.uk/OrderDetails.do?uin=uk.bl.ethos.775857}

\bibitem{doct5query}
Nogueira, R., Lin, J.: From doc2query to doctttttquery. In: Online preprint 6
  (2019), \url{https://github.com/castorini/docTTTTTquery}

\bibitem{noor2015evaluating}
Noor, S., Bashir, S.: Evaluating bias in retrieval systems for recall oriented
  documents retrieval. International Arab Journal of Information Technology
  (IAJIT)  \textbf{12}(1) (2015)

\bibitem{palma}
Palma, J.G.: Homogeneous middles vs. heterogeneous tails, and the end of the
  ‘inverted-u’: the share of the rich is what it's all about. Cambridge
  working papers in economics, Faculty of Economics, University of Cambridge
  (2011), \url{https://EconPapers.repec.org/RePEc:cam:camdae:1111}

\bibitem{aol-log}
Pass, G., Chowdhury, A., Torgeson, C.: A picture of search. In: Proceedings of
  the 1st International Conference on Scalable Information Systems. p. 1–es.
  InfoScale '06, Association for Computing Machinery (2006).
  \doi{10.1145/1146847.1146848}

\bibitem{pickens2010reverted}
Pickens, J., Cooper, M., Golovchinsky, G.: Reverted indexing for feedback and
  expansion. In: Proceedings of the 19th ACM international conference on
  Information and knowledge management. pp. 1049--1058 (2010)

\bibitem{roy2022studying}
Roy, D., Carevic, Z., Mayr, P.: Studying retrievability of publications and
  datasets in an integrated retrieval system. In: Proceedings of the 22nd
  ACM/IEEE Joint Conference on Digital Libraries. JCDL '22, Association for
  Computing Machinery (2022). \doi{10.1145/3529372.3530931}

\bibitem{roy2023retrievability}
Roy, D., Carevic, Z., Mayr, P.: Retrievability in an integrated retrieval
  system: an extended study. International Journal on Digital Libraries  (Apr
  2023). \doi{10.1007/s00799-023-00363-4},
  \url{https://doi.org/10.1007/s00799-023-00363-4}

\bibitem{traub2018impact}
Traub, M.C., Samar, T., van Ossenbruggen, J., Hardman, L.: Impact of
  crowdsourcing ocr improvements on retrievability bias. In: Proceedings of the
  18th ACM/IEEE on Joint Conference on Digital Libraries. p. 29–36. JCDL '18,
  Association for Computing Machinery (2018). \doi{10.1145/3197026.3197046}

\bibitem{traub2016querylog}
Traub, M.C., Samar, T., Van~Ossenbruggen, J., He, J., de~Vries, A., Hardman,
  L.: Querylog-based assessment of retrievability bias in a large newspaper
  corpus. In: 2016 IEEE/ACM Joint Conference on Digital Libraries (JCDL). pp.
  7--16. IEEE (2016)

\bibitem{overviewRobust}
Voorhees, E.M.: Overview of the {TREC} 2004 robust track. In: Voorhees, E.M.,
  Buckland, L.P. (eds.) Proceedings of the Thirteenth Text REtrieval
  Conference, {TREC} 2004, Gaithersburg, Maryland, USA, November 16-19, 2004.
  {NIST} Special Publication, vol. 500-261. National Institute of Standards and
  Technology {(NIST)} (2004),
  \url{http://trec.nist.gov/pubs/trec13/papers/ROBUST.OVERVIEW.pdf}

\bibitem{wilkie2013initial}
Wilkie, C., Azzopardi, L.: An initial investigation on the relationship between
  usage and findability. In: Advances in Information Retrieval: 35th European
  Conference on IR Research, ECIR 2013, Moscow, Russia, March 24-27, 2013.
  Proceedings 35. pp. 808--811. Springer (2013)

\bibitem{wilkie2013relating}
Wilkie, C., Azzopardi, L.: Relating retrievability, performance and length. In:
  Proceedings of the 36th international ACM SIGIR conference on Research and
  development in information retrieval. pp. 937--940 (2013)

\bibitem{wilkie_2014}
Wilkie, C., Azzopardi, L.: Best and fairest: An empirical analysis of retrieval
  system bias. In: de~Rijke, M., Kenter, T., de~Vries, A.P., Zhai, C., de~Jong,
  F., Radinsky, K., Hofmann, K. (eds.) Advances in Information Retrieval. pp.
  13--25. Springer International Publishing, Cham (2014)

\bibitem{wilkie2014efficiently}
Wilkie, C., Azzopardi, L.: Efficiently estimating retrievability bias. In:
  de~Rijke, M., Kenter, T., de~Vries, A.P., Zhai, C., de~Jong, F., Radinsky,
  K., Hofmann, K. (eds.) Advances in Information Retrieval. pp. 720--726.
  Springer International Publishing, Cham (2014)

\bibitem{wilkie2014retrievability}
Wilkie, C., Azzopardi, L.: A retrievability analysis: Exploring the
  relationship between retrieval bias and retrieval performance. In:
  Proceedings of the 23rd ACM International Conference on Conference on
  Information and Knowledge Management. pp. 81--90 (2014)

\bibitem{wilkie2015queryLength}
Wilkie, C., Azzopardi, L.: Query length, retrievability bias and performance.
  In: Proceedings of the 24th ACM International on Conference on Information
  and Knowledge Management. p. 1787–1790. CIKM '15, Association for Computing
  Machinery, New York, NY, USA (2015). \doi{10.1145/2806416.2806604},
  \url{https://doi.org/10.1145/2806416.2806604}

\bibitem{wilkie2015retrievability}
Wilkie, C., Azzopardi, L.: Retrievability and retrieval bias: A comparison of
  inequality measures. In: Advances in Information Retrieval: 37th European
  Conference on IR Research, ECIR 2015, Vienna, Austria, March 29-April 2,
  2015. Proceedings 37. pp. 209--214. Springer (2015)

\bibitem{zheng2009document}
Zheng, L., Cox, I.J.: Document-oriented pruning of the inverted index in
  information retrieval systems. In: 2009 International Conference on Advanced
  Information Networking and Applications Workshops. pp. 697--702. IEEE (2009)

\bibitem{ahu2021aol}
Zhu, Y., Nie, J.Y., Dou, Z., Ma, Z., Zhang, X., Du, P., Zuo, X., Jiang, H.:
  Contrastive learning of user behavior sequence for context-aware document
  ranking. In: Proceedings of the 30th ACM International Conference on
  Information \& Knowledge Management. p. 2780–2791. CIKM '21, Association
  for Computing Machinery (2021). \doi{10.1145/3459637.3482243}

\end{thebibliography}
\end{document}